# Possibility for Control and Optical Filter Wheel Positioning Based on a Hall Sensor

*At. Atanassov, L. Bankov*

*Solar-Terrestrial Influences Laboratory, Bulgarian Academy of Sciences, Stara Zagora Division, Bulgaria,*
*At_M_Atanassov@yahoo.comy*

*In memoriam of our colleague*
*Krassimir Kanev*

*This paper presents a controlling and positioning approach for an optical filter disk. In the known applications where Hall-effect sensors and magnets (radially positioned) are applied, the purpose of the latter is to help developing a binary code used mainly for filter identification. The number of sensors is equal to the maximum number of magnets that participate in forming the filter identity code. Therefore, three sensors and the corresponding number of magnets radially positioned for each position are required for the identification of seven filters. The precise position of the filter disk is reached by counting up the steps of a stepper motor.*

*The presented approach applies only one Hall-effect sensor. The fore working front of one magnet, positioned along the disk periphery, is used to precisely mark each filter position. At a certain distance after the main magnet, a second one is positioned in order to form an identity code.*

*The paper contains an algorithm for optical filter disk control and positioning, along with the mathematical formalism on which it is based.*

## Introduction

The present status of optical measurements carried out in different fields of the fundamental and applied science is based on the employment of optical filter sets. A possibility to rapidly change the filters' location in the course of an observation session is to arrange the filters on a rotating wheel. The precise rotation speed and the reliable positioning are crucial for the measurements in course. Nowadays, when the optic instruments are a part of computer measurement systems, the filter wheel rotation is controlled by a specific controller. Different options for allocating the tasks between the PC and the controller are now available. The former entirely circuit-based solutions of the controllers are replaced by the application of programmable micro- or PIC controllers.

The precise positioning of the filter wheel system is critical. While opto-couples were used in the past to mark the position, now Hall-effect sensors with magnets are applied. Filter wheels equipped with twelve filters are incorporated into a WIRO Prime as sets composed of one to three magnets positioned in four standard positions against each filter are detecting the precise position. This approach is analogical to the application of opto-couples [1]. Four Hall-effect sensors are required to read out the available magnets. Three-bit coding is applied for filter identification (similarly to the previous case, where the code is a four-bit one) in Omega2000 IR camera [2]. This is implemented by sets of one to three magnets radially positioned on a filter wheel and three Hall-effect sensors to detect them, respectively.

The Gemini Near-Infrared Integral-Field Spectrograph (NIFS) employs one Hall-effect sensor and one magnet for every filter position in the course of a unidirectional rotation [3]. The middle of every magnet is found for position determination - the place between the two fronts where the sensor begins to respond to the magnetic field.

Initially, the Spectrometer Airglow Temperature Imager (SATI) [5, 6] incorporates a four-filter nests wheel and a nine opto-couples system positioned on the pivot shaft.

The SATI was subsequently improved by Stara Zagora Division of the Solar-Terrestrial Influences Laboratory [7] within the framework of several contracts entered into with the Centre for Space Research in Earth and Space Science (CRESS) of the York University in Toronto, Canada. A part of this work was focused on the development of a positioning system and a control model [8].

After the first positioning system was developed and applied by two SATI instruments manufactured in 2001 for the Korea Ocean Research and Development Institute and located respectively on Resolute Bay (74.07°N, 265.01°E), Canada, and King Sejong Antarctica (42.43°N, 25.62°E), and in 2003 for Solar-Terrestrial Influences Laboratory (42.43°N, 25.62°E), CRESS assigned to the Solar Terrestrial Influences Laboratory the development of a filter wheel of six and more filter positions.

This paper present an optical filter wheel control and positioning model, developed and applied in SATI instrument, one for Kazakhstan and one for China, respectively, and several for Canada.

## General Characteristic of the First Position System Option

The first positioning system option applies one magnet only for each filter position. An additional magnet identifies some of the filters (positions) as a first position. This is also achievable without applying an additional magnet, if the dimension of some of the magnets is selected to be different from the others. At the beginning of every operation of the filter wheel an initialization is carried out during which the wheel performs a full rotation to accumulate information of the relative positions of all magnets. The information about the positions of all magnets, which is retrieved by the Hall-effect sensor, is analyzed and the wheel is set at one initial position.

The searching of the targeted position is related to identifying the work front of the magnets. When a disturbance occurs during the control, a work front can be detect for a position, different from the one determined during the initialization. Under this circumstance, the initialization is repeated and the search is restarted.

This control and positioning option is the most simplified and reliable one both in terms of manufacturing technology and control. It is very important to have the initialization process free from any disturbances. In this particular case, such disturbances are related to the fact that an initialization is performed of a step of the stepper motor but no such a step is actually performed due to an unknown reason. Such disturbances are very rare, especially during nocturnal measurement sessions. Furthermore, the control algorithm itself tolerates the presence of such disturbance to some extent. However, the analysis shows that, if such disturbance occurs during the initialization, though being very rare, it is likely that the filter wheel control detects incorrect positions.

## General Description of SATI Mechanical Sub-System

The filter wheel is driven by a belt transmission stepper motor (Fig.1). The motor step is 7,5°. The number of steps for one full rotation of the wheel, by suitable reduction gear, is 960. The filter wheel radius is 20 cm. The cogs along the wheel periphery serve to transmit motion. Experimentally one cogless wheel was manufactured, which is also operating.

Depending on the physics problems to be solved, the number of filters that can be placed on the wheel may be different. It is determined both by the wheel dimensions and the filters dimensions. Two filter wheel options were manufactured with 4 and 6 filter nests, respectively.

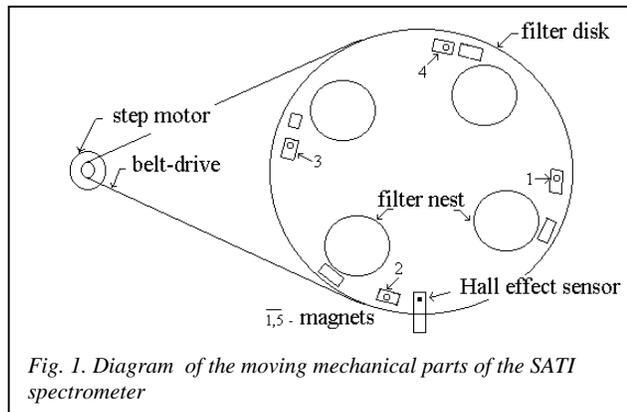

*Fig. 1. Diagram of the moving mechanical parts of the SATI spectrometer*

For more filters being requested and for reliability reasons in the filter control and positioning, a new positioning system option was launch. One Hall-effect sensor (similarly to the previous option) and a set of magnet couples positioned along the filter wheel periphery are applied in this option to form an identity code.

## Features of the Suggested Positioning System

The magnetic front (MF) is the milestone of the control strategy. The MF front is the place where the Hall-effect sensor starts registering the magnetic field, while passing over the magnet. The filter position is not identified by the whole magnet, but by only one of its two fronts - a working MF.

The experiments found out that the front of the magnet field represents a very narrow linear section where magnetic field begins to act on Hall-effect sensor by its linear shift. The counting of Hall-effect sensor signals, having a frequency of 100 times bigger than the frequency of the stepper motor, shows an exclusive stability over time of the event related to passing over MF. Therefore, MF is able to secure a precision of the positioning, better than 1% from the angle shift corresponding to one step of the stepper motor within the applied mechanical system. This would require a stepper motor and a ratio of the transmission gear that secures a much less linear shifting and filter wheel rotation.

The electronic circuit of the stepper motor control rotates the filter wheel in two directions. The forwards rotational direction is the one by the means of which the Hall-effect sensor meets the working front first. In the reverse rotation direction, the Hall-effect sensor meets the second PS first. This serves to only detect the magnet rather than making the positioning more precise. A very simple solution of this problem could be related to the passing of the working front and the reverse motion for its detection. This takes one additional step only and does not cause any algorithmical problems.

Unlike the positioning system described in [8], the present one uses the fore working front of one magnet to point the right positioning place of each filter, while the second magnet, situated at a certain distance behind the first one, is used to identify the position (Fig. 1).

The logical diagram depicting the work of the positioning system is shown on Fig. 2. In order to avoid cluttering, only four positions corresponding to four filter nests- $f_1$, $f_2$, $f_3$ and $f_4$ are shown. Every position is determined by a couple of magnets- (1,1'), (2,2'), (3,3') and (4,4'). One of the two magnets is taken as basic, connected to the precise determination of the filter position. The second magnets, marked with the "'" symbol, are auxiliary and serve to form the identity codes of every filter. Each basic magnet has a fore or working (w) and a back (h) front. There are the locations where magnetic field induction is enough to influence on the Hall-effect sensor. The outer front of the couple of the basic magnet is considered to be the working front (w). A precise tuning is required so as to enable the working front performing its function.

## Objectives and Tasks of the Filter Wheel Control Design

The design and development of the filter wheel positioning system and control is focused on the following objectives [I]:
– improvement of the positioning precision;
– decreasing the positioning time;
– reliability in outdoor operation;
– reducing the probability of errors.

The additional objectives were focused on articulating ideas and solutions to achieve constructive and technological simplification in the manufacturing process. The objectives strived for during the development of the first option are still valid.

## Mathematical Formalism to Back up the Positioning Problems

The positioning system (PS) can be treated as a combination of two synchronized finite automata The first automatum $F_1(Z_1, X_1, Y_1, \varphi_1, \psi_1, z_1^0)$ is determined by a 6-tuple $\langle Z_1, X_1, Y_1, \varphi_1, \psi_1, z_1^0 \rangle$, where $Z_1$ is multitude of internal conditions, $X_1$- multitude of input signals (filters), $Y_1$- output signals, $\varphi_1$- transition function $(z(t+dt) = \varphi_1(z(t), x(t)))$, $\psi_1$- output function $(y(t+dt) = \psi_1(z(t), x(t)))$ and $z_1^0$- initial condition. Similarly, as regards the second automatum, it may be written: $F_2(Z_2, X_2, B, Y_2, \varphi_2, \psi_2, z_2^0; \alpha = Y_1)$. The binary multitude B contains input signals which reflect the signatures of external environment; in our case, the signatures of the environment are determined based on the magnetic field of the applied magnets. A parameter $\alpha$ determines the working modes of $F_2$, which is equivalent to the pointing of some sub-multitude of $Z_2$.

The automatum $F_1$ determines an optimal transition related to the selection of the direction of filter wheel rotation, i.e. finding the shortest way to the targeted position. The output signal of $F_1$ - $y_1(t)$ is an additional parameter which determines the working mode of $F_2$ by the possible sequential transitions.

Unlike [8], where automatum $F_2$ searches working MF only for the serial magnet, in this case it searches a binary image that identifies the target magnet. The multitude X2 contains the input signals, and determines searching pattern in the second input multitude B. The B contains a description of the external environment where the searching is performed. B is a binary infinite multitude whose elements are signals measured by the Hall-effect sensor.

Depending on the length of the searched image, such a pattern is established by comparisons, containing input signals from previous times for each time of automaton $F_2$. Therefore, $F_2$, in its substance, presents an option of a secondary order Mealy automatum

$$\begin{cases} z_2(t_{k+1}) = \varphi_2[z_2(t), x_2(t), y_1(t_n)] \\ y_2(t_k) = \psi_2[z_2(t), b_{t_k}, b_{t_{k-1}}, ..., b_{t_{k-m-1}}] \end{cases}$$

For the multitude B containing input signals related to the magnetic field registered by the Hall-effect sensor, which are input in a computer, there could be interferences demonstrated in this particular case by loss of control impulses to the stepper motor, where the Hall-effect sensor signals are always input. It is equivalent to introducing extra signals $b_{t_m} = b_{t_{m-1}}$. Therefore, the searching approach is replaced by recognition.

The searching of every targeted position **m** $(m = \overline{1, M})$ is carried out by binary patterns $T_m$ with a length of $L_m$. Every pattern $T_m$ contains an identity code for the respective positions. These patterns are launched when the wheel is manufactured and are tuned up for every position. Patterns for forwards filter wheel direction are prepared. If a backwards rotation, inverted patterns are applied

$$\overline{T}_m(l) = T_m(L_m - l + 1).$$

After the identity code is established, a backwards motion is required, so as to set the filter wheel at the targeted position, determined by the working front of the positioning magnet.

The searching of the filter position identification pattern looks like searching a word in a text. There are algorithms of various complexities and effectiveness [9]. In our case, the direct search would be enough and fully compliant with the requirement for an algorithm that is relatively elementary and fast enough.

## Algorithm for Rotation Optimization During the Positioning

As already mentioned, $F_1$ selects the rotation direction of the filter wheel. This is to reduce as much as possible the angle of rotation, and therefore, the time to reaching to the targeted position. Thus, unlike any other filter wheel elaborations, the maximal rotation for achieving the most distant position is within half a revolution.

## Filter Wheel Control Commands

Filter wheel motion and positioning command system is offered to improve efficiency, flexibility and convenience when composing the measurement control cyclograms. They are developed by analogy with the control commands of other photometric systems [8]. The following procedures are developed in fortran95 language on Digital fortran compiler:

- **Filter_wheel_initialization** - this command is executed only once before the first execution of the positioning command. It is to read the information about filter positions patterns. Subsequently, the filter wheel takes one of the possible positions.
- **Set_Filter**(filter_number) - this is the basic function for selecting and positioning the filter with a serial number **filter_number**.

In terms of achieving simplicity and convenience in patterns determination, the patterns' length is selected to be a half of the distance between two adjacent filter positions expressed in unit steps of the stepper motor

$$L \leq \frac{N/m}{2},$$

where N are the steps for one full filter wheel revolution and m is the number of the filter positions. In case of 960 steps and six filters, the maximal length of the patterns is 80 steps. We may have patterns of different length if only the distance between the fore working front of the basic magnet and the back front of the second one is applied. So, by using two magnets having a length of 3-4 steps and a distance between them of 3-4 steps, the length of the end pattern will be about 10 steps.

## Precision and Positioning Time

The positioning precision is achieved by applying a working front only (which is very narrow) for marking the filter positions and it is about 22′ - half of the angular step. The decrease of the step of the stepper motor enhances the positioning precision. Disturbances in the control electronics (especially if the distance between the control PC with the controller and the optical instrument which includes the filter wheel is long) may cause some steps initiated by the stepper motor remaining incompleted. The control algorithm compensates entirely this problem.

The filter wheel we have developed is bidirectional, in contrast to the known ones, described in [1, 2, 3, 4]. The time taken by one arbitrary filter to move to the opposite one does not exceed 10sec for all the filter wheels manufactured by us. No more than half a revolution of the filter wheel is necessary for the arbitrary change of one filter position to another.

The transition time between two adjacent positions is about 3 sec when rotation is backwards and 6 sec when the rotation is forwards and under equal length patterns. If different length patterns and configuration than the ones shown on Fig.2 are used by moving backwards, the pattern length does not sign on position times. However, by moving forwards, the transitions begin from the working front of magnet on initial position. The rotation continues to the back front of the secondary magnet of the target filter, whereupon is necessary reverse to working front of the basic magnet.

## Conclusion

The presented positioning approach and control model are flexible in terms of different parameters of the mechanical sub-system, i.e. the number of filters, steps of the stepper motor. The code portability is connected to the employment of the highly standardized program language Fortran, used for coding purposes.

In [4] realizations, applying Hall-effect sensors, the magnets are primarily used to identify the target filter position, where the precise position is reached by counting up a particular number of steps of the stepper motor. In the presented approach, the number of steps is not important, for a working front of the magnet is applied to properly position the filter. This may be advantageous in environments with interferences, where the motion may go on until the proper position of the filter is identified.

In the presented positioning system, the number of filter positions is dependent on neither the number of the applied sensors nor the number of magnets. However, the number of magnets M used to form an identity code may be more than two (M >2). It is even possible to use sets composed of various number of magnets for every position. It would be interested to investigate the relation between the number of magnets and their sizes so as to form an error resistant code.

The fore working front of the magnets is very narrow and allows, in a very tiny step of the motor and a good setting, achieving the targeted level of positioning.

**Acknowledgements:** The authors would like to thank Dr. B. H. Solheim and Dr. S. I. Sargoytchev from the Centre for Research in Earth and Space Science (CRESS) of the York University in Toronto, Canada, for their help and advice concerning the SATI instrument and the data acquisition. The authors would like to thank Mrs. M. Nikolova for the technical assistance.

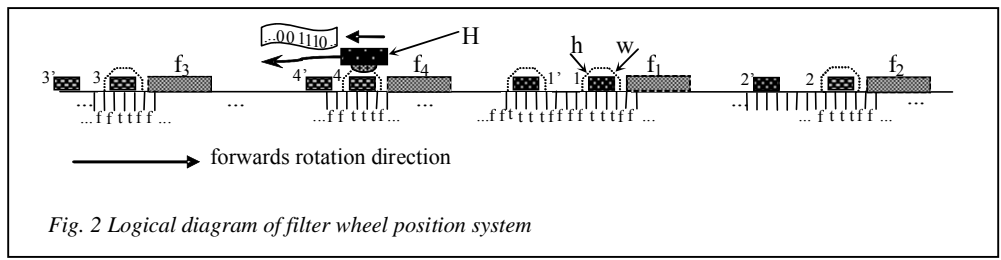

*Fig. 2 Logical diagram of filter wheel position system*